\documentclass[11pt,fullpage]{article}

\usepackage[utf8]{inputenc} % allow utf-8 input
\usepackage[T1]{fontenc}    % use 8-bit T1 fonts
\usepackage{hyperref}       % hyperlinks
\usepackage{url}            % simple URL typesetting
\usepackage{booktabs}       % professional-quality tables
\usepackage{amsfonts}       % blackboard math symbols
\usepackage{nicefrac}       % compact symbols for 1/2, etc.
\usepackage{microtype}      % microtypography
\usepackage{amsmath}
\usepackage{graphicx}

\DeclareMathOperator*{\argmin}{arg\,min}
\usepackage{authblk}

\title{Extreme Few-view CT Reconstruction using Deep Inference}

\author{Hyojin Kim\footnote{Corresponding Author: hkim@llnl.gov}, Rushil Anirudh, K. Aditya Mohan, Kyle Champley}
\affil[]{Lawrence Livermore National Laboratory, Livermore, California.}
\date{}

\begin{document}

\maketitle

\begin{abstract}
  Reconstruction of few-view x-ray Computed Tomography (CT) data is a highly ill-posed problem. It is often used in applications that require low radiation dose in clinical CT, rapid industrial scanning, or fixed-gantry CT. Existing analytic or iterative algorithms generally produce poorly reconstructed images, severely deteriorated by artifacts and noise, especially when the number of x-ray projections is considerably low. This paper presents a deep network-driven approach to address extreme few-view CT by incorporating convolutional neural network-based inference into state-of-the-art iterative reconstruction. The proposed method interprets few-view sinogram data using attention-based deep networks to infer the reconstructed image. The predicted image is then used as prior knowledge in the iterative algorithm for final reconstruction. We demonstrate effectiveness of the proposed approach by performing reconstruction experiments on a chest CT dataset. 
\end{abstract}

\let\thefootnote\relax \footnotetext{This work was performed under the auspices of the U.S. Department of Energy by Lawrence Livermore National Laboratory under Contract DE-AC52-07NA27344. LLNL-CONF-791857}

\section{Introduction}

Computed Tomography (CT) reconstruction is an inverse problem where images are reconstructed from a collection of multiple x-ray projections known as sinogram. Conventional CT imaging systems use densely sampled x-ray projections (roughly equal to one projection per detector column) with a full angular range (180-360 degrees). Unlike the conventional CT setup, on the other hand, some CT systems use different imaging configurations that require rapid scanning or reduced radiation dose. In those cases, the CT imaging uses a small number of x-ray projections, referred to as few-view CT. Reconstructing images from a few x-ray projections becomes an extremely under-determined inverse problem, which results in significant image degradation. The reconstructed images from extremely few-view sinogram measurement (10 views or less) are often characterized by severe artifacts and noise, even with state-of-the-art regularized iterative algorithms ~\cite{Fessler_TIP_1999, Herman2008, Sidky_PMB_2008, Ritschl2011, Xu2012} as well as with the widely used Filtered Backprojection (FBP) ~\cite{Kak2001}. 

In recent years, deep learning-based approaches have been successfully applied to a number of image restoration, denoising, inpainting and other image processing applications. Methods in this category use perceptual information as well as contextual features to improve the image quality. In CT imaging applications, several deep convolutional neural network (CNN) approaches have been proposed to address different ill-conditioned CT reconstruction applications. Methods in ~\cite{Chen2017, Jin2017, Ye2018} proposed CNN-based approaches to improve the image quality for low-dose (sparse-view) imaging. These approaches aim to infer the noise distribution to generate a cleaner image from the noisy image. However, these methods do not employ the sinogram to ensure that the reconstructed image is consistent with the measurement. Gupta et al. ~\cite{Gupta2018} proposed a method using a CNN-based projector for moderate sparse-view reconstruction (45 and 144 views). Anirudh et al. ~\cite{Anirudh2018} proposed a CNN-based sinogram completion approach to address limited-angle CT reconstruction. 

% Chen et al. \ref{Chen2017} proposed a residual encoder-decoder algorithm to improve the image quality for low-dose CT imaging. 
% Jin2017 proposed an U-net-adopted CNN approach to improve the image quality.
% Ye2018 for few-view using single-view back projection images. 

In this paper, we present a CNN inference-based reconstruction algorithm to address extremely few-view CT imaging scenarios. For the initial reconstruction, we employ a CNN-based inference model, based on CT-Net ~\cite{Anirudh2018}, that directly uses the input measurement (few-view sinogram data) to predict the reconstructed image. In the cases where the sinogram measurements are extremely under-sampled, the images reconstructed by existing analytic and iterative methods may suffer from too much noise with little high frequency information, and the methods in ~\cite{Chen2017, Jin2017, Ye2018} may repair the missing or noisy part with perceptually created, but incorrect content. Thus, we pursue a method that directly uses the sinogram so that the reconstructed content is consistent with the input measurement, as an inverse problem. Furthermore, instead of performing the sinogram completion in ~\cite{Anirudh2018} optimized for limited-angle reconstruction, we propose to use the predicted image from the CNN inference model as an image prior in state-of-the-art iterative algorithms in order to improve the final reconstruction. Our experiments on a chest CT dataset show that the proposed model outperforms existing analytical and state-of-the-art iterative algorithms as well as the sinogram completion.

\begin{figure*}
\begin{minipage}[c]{1\linewidth}
\centering
\includegraphics[width = 0.95\linewidth]{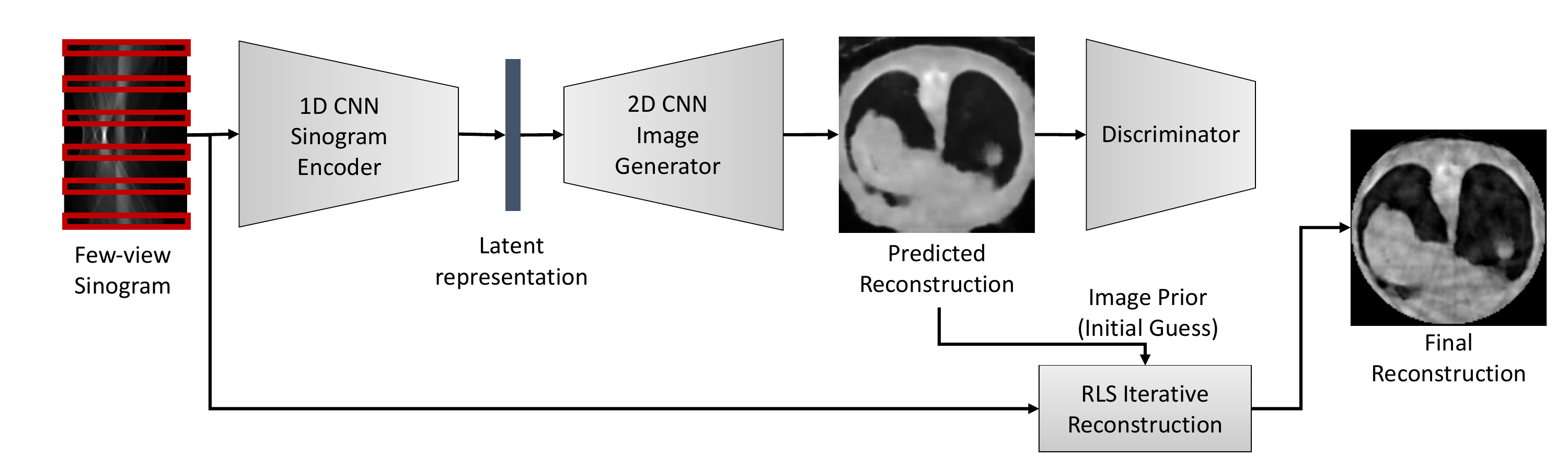}
\end{minipage}
\caption{An overview of the proposed approach for extreme few-view CT reconstruction.}
\label{fig:overview}
\end{figure*}

\section{Proposed Method}

The proposed approach is comprised of two stages: (a) CNN-driven inference, (b) iterative reconstruction with the image prior, as illustrated in Figure \ref{fig:overview}. The algorithm begins with CNN-based deep inferencing to predict the reconstructed images from few-view sinogram data. Then we perform an iterative algorithm with the predicted images as initial guess. 

For the CNN-based deep inferencing, we adopt CT-Net proposed in \cite{Anirudh2018} but modify it for few-view CT applications. Our network architecture is similar to the original CT-Net, consisting of 1D-2D CNNs to infer images directly from the sinogram data. Unlike limited-angle reconstruction, however, the sinogram in few-view CT is spatially less coherent and the number of projections is smaller compared to limited-angle reconstruction, and we use smaller filter sizes in the 1D CNN. For the training loss, we use SSIM ~\cite{Wang2004} loss in addition to $L_{2}$ and adversarial loss with a discriminator to generate more realistic CT images. We empirically choose the weight between $L_{2}$ and SSIM losses (0.7 in our experiment).

The predicted reconstructed images are then used as image prior in the Regularized Least Squares (RLS)-based iterative algorithm ~\cite{Fessler_TIP_1999, Ritschl2011, Xu2012}. In RLS, images are reconstructed by solving the following optimization:
\begin{equation}
\label{eq:RLS}
\hat{X} = \argmin_{x} { \|y - Ax\|^2 + \beta R(x)}
\end{equation} where $y$ is measured sinogram data, $\hat{X}$ is the reconstructed attenuation coefficients of the object, $A$ is the x-ray projection matrix subject to the projection geometry, and $R(x)$ is a Total Variation regularization functional, and $\beta$ serves as a weight to control the strength of the regularization. The $\beta$ term is determined empirically. 

The optimization typically requires $50-100$ iterations, starting with a randomly chosen initial guess. However, the optimization with a carefully selected initial guess enables high quality reconstruction as well as fast convergence. To this end, we propose to use our deep network-driven prediction as image prior (initial guess) in this iterative algorithm, which enables to recover edge-preserving high frequency regions in the reconstructed images. There are a set of parameters to be chosen empirically in the RLS algorithm. For our experiment on the chest CT dataset, we chose $100$ iterations with $\beta=2e^{-2}$ in \eqref{eq:RLS} and the non-negative constraint. 
% without falling into a local minimum

\begin{figure*}
\centering
\includegraphics[width=0.97\linewidth]{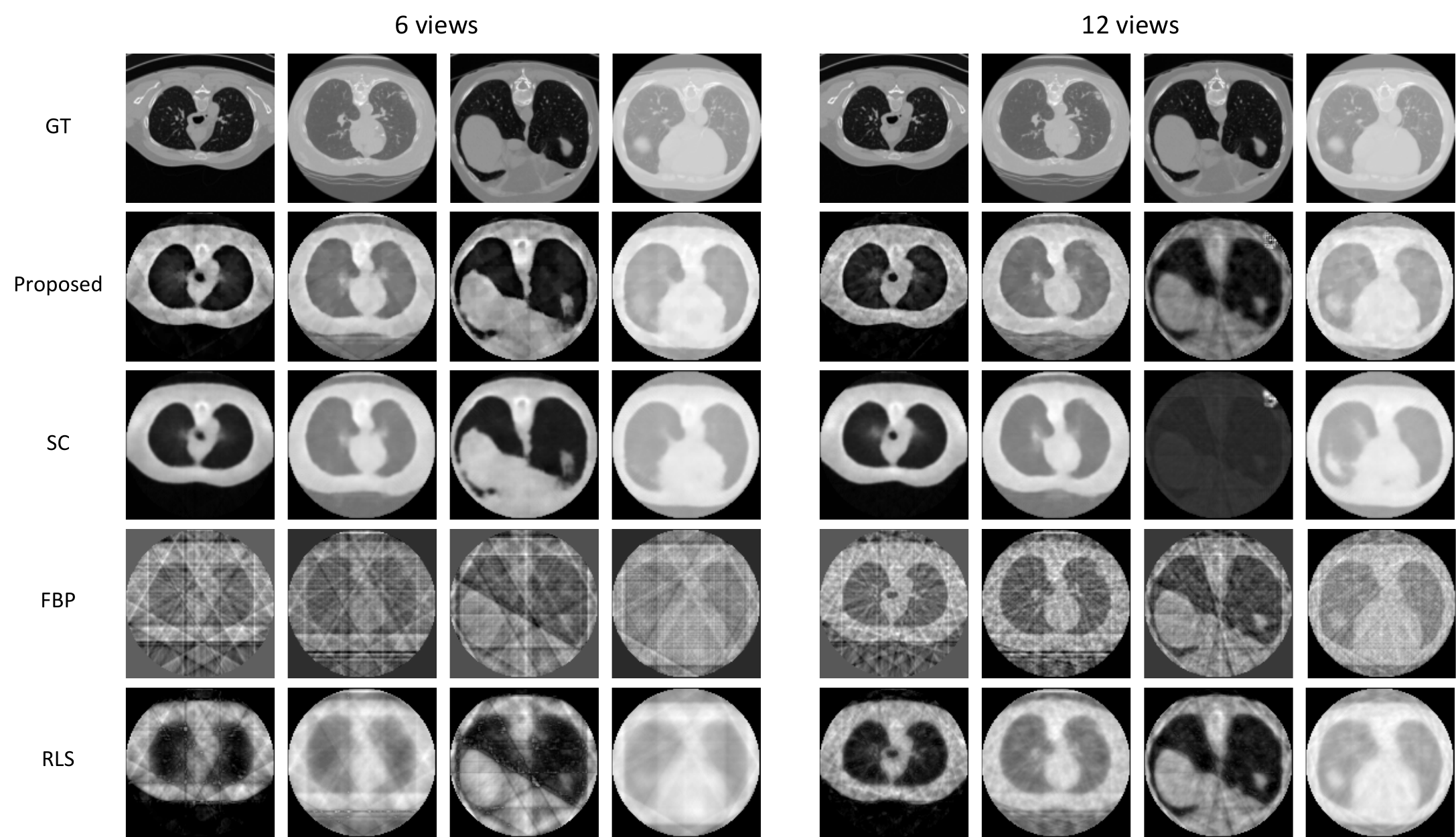}
\caption{Reconstruction results from 6 and 12 views.}
\label{fig:result_img}
\end{figure*}

\section{Experiments}

To demonstrate effectiveness of the proposed approach, we perform experiments on a chest CT dataset provided by the National Cancer Institute ~\cite{NCI_Lung}. The chest CT dataset is comprised of 1595 patients of chest scans for the study of lung nodule detection as an early stage cancer diagnosis. We split the entire scans into 1,000 patients of scans for training and the rest for testing. The reconstructed volumes are $512 \times 512 \times N$ where $N$ is the number of slices ranging from 94 to 541, and each slice is treated independently to demonstrate the parallel-beam projection based reconstruction. We down-sampled the original $512 \times 512$ slices into $128 \times 128$ to fit the proposed model to a single GPU memory. Since the dataset does not provide the original sinogram, we performed forward projection using Livermore Tomography Tools ~\cite{LTT} to generate sinogram of $180 \times 128$ where 180 is the number of projection views. We then sampled sinogram for the training and test input to perform few-view experiments. In 9-view reconstruction, for example, we sampled views at 0, 20, 40, 60, 80, 100, 120, 140, 160 degrees from 180 views. 

For the training, we used Adam optimizer with learning rate of $1 \times 10^{-3}$ and the exponential decay of every 10,000 steps with a base of 0.97. The mini-batch size is 50 and the number of epochs is 50. We trained our model using Tensorflow on NVIDIA GeForce RTX 2080 Ti. To evaluate the reconstruction performance, we randomly sampled 50 images from the test dataset and performed our reconstruction together with FBP, RLS and Sinogram Completion (SC) ~\cite{Anirudh2018} as baseline algorithms. We report PSNR and SSIM of the reconstruction results. 

\begin{figure*}
%\begin{minipage}[c]{1\linewidth}
\centering
\includegraphics[width=0.49\linewidth]{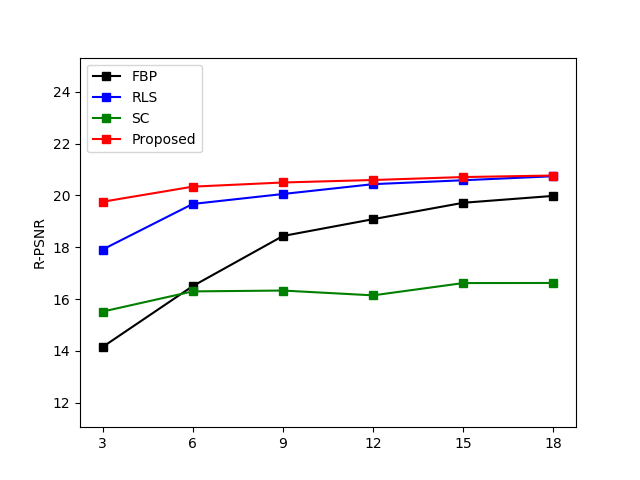}
\includegraphics[width=0.49\linewidth]{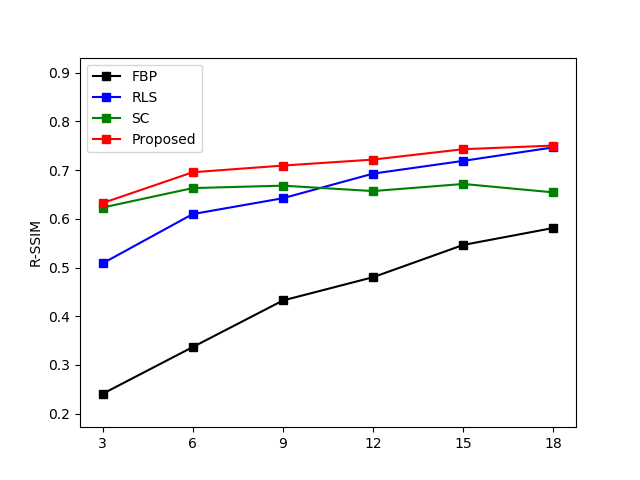}\\
\includegraphics[width=0.9\linewidth]{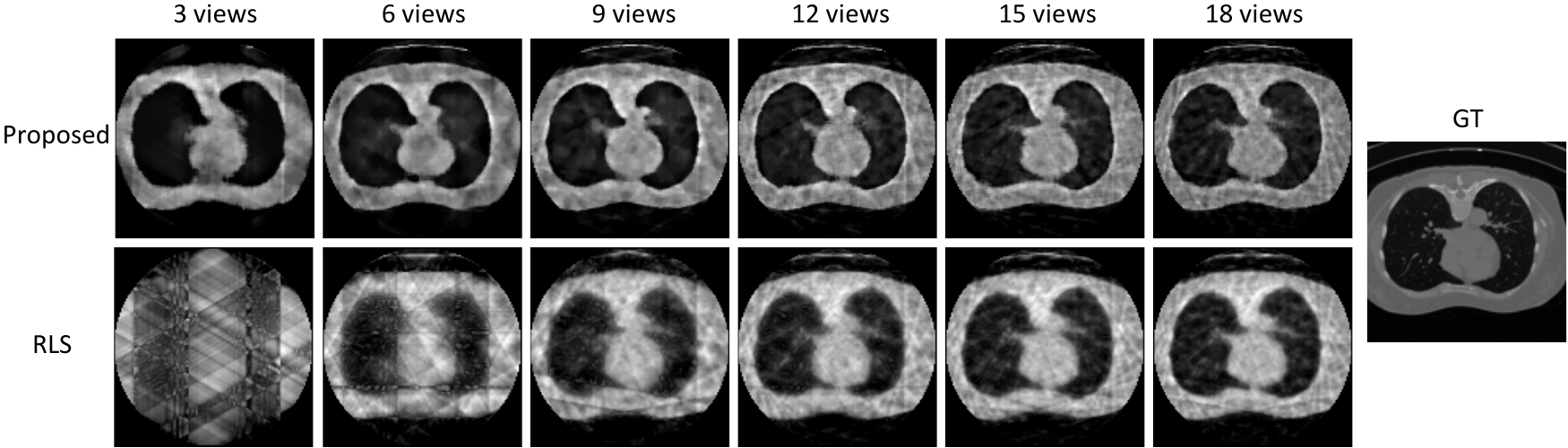}\\
%\end{minipage}
\caption{quantitative evaluation from different views. PSNR (\textit{top left}) and SSIM (\textit{top right}) as an image quality metric on 50 test samples. The bottom images are results of our proposed method and RLS from different views.}
\label{fig:result_plot}
\end{figure*}

Figure \ref{fig:result_img} shows qualitative results from two different views, 6 and 12, respectively. We observe that the proposed method yields sharper edges and more high frequency regions with less artifact and noise, compared to the baseline algorithms. Figure \ref{fig:result_plot} shows quantitative evaluation on 6 different few-view experiments (3, 6, 9, 12, 15, 18 views). PSNR and SSIM show that the proposed approach outperforms the baselines although both metrics are not sufficient to measure perceptual quality. Another observation is that RLS improves the image quality as the number of views increases while our method outperforms by a large margin especially when the number of views is extremely small.

\small{\subsubsection*{Disclaimer}
\noindent This document was prepared as an account of work sponsored by an agency of the United States government. Neither the United States government nor Lawrence Livermore National Security, LLC, nor any of their employees makes any warranty, expressed or implied, or assumes any legal liability or responsibility for the accuracy, completeness, or usefulness of any information, apparatus, product, or process disclosed, or represents that its use would not infringe privately owned rights. Reference herein to any specific commercial product, process, or service by trade name, trademark, manufacturer, or otherwise does not necessarily constitute or imply its endorsement, recommendation, or favoring by the United States government or Lawrence Livermore National Security, LLC. The views and opinions of authors expressed herein do not necessarily state or reflect those of the United States government or Lawrence Livermore National Security, LLC, and shall not be used for advertising or product endorsement purposes.
}

{\small
\bibliographystyle{unsrt}  
\bibliography{egbib}

\begin{thebibliography}{10}

\bibitem{Fessler_TIP_1999}
Jeffrey~A. Fessler and Scott~D. Booth.
\newblock Conjugate-gradient preconditioning methods for shift-variant pet
  image reconstruction.
\newblock {\em IEEE Trans. Imag. Proc.}, 8:688--699, 1999.

\bibitem{Herman2008}
G.~T. Herman and R.~Davidi.
\newblock Image reconstruction from a small number of projections.
\newblock {\em Inverse Problems}, 24(4), 2008.

\bibitem{Sidky_PMB_2008}
E.~Y. Sidky and X.~Pan.
\newblock Image reconstruction in circular cone-beam computed tomography by
  constrained, total-variation minimization.
\newblock {\em Phys. Med. Biol.}, 53:4777--4807, 2008.

\bibitem{Ritschl2011}
L.~Ritschl, F.~Bergner, C.~Fleischmann, and M.~Kachelrieß.
\newblock Improved total variation-based ct image reconstruction applied to
  clinical data.
\newblock {\em Physics in Medicine and Biology}, 56(6):1545--–1561, 2011.

\bibitem{Xu2012}
Q.~Xu, H.~Yu, X.~Mou, L.~Zhang, J.~Hsieh, and G.~Wang.
\newblock Low-dose x-ray ct reconstruction via dictionary learning.
\newblock {\em IEEE Transactions on Medical Imaging}, 31(9):1682--–1697,
  2012.

\bibitem{Kak2001}
A.~C. Kak and M.~Slaney.
\newblock {\em Principles of Computerized Tomographic Imaging}.
\newblock Society of Industrial and Applied Mathematics, 2001.

\bibitem{Chen2017}
H.~Chen, Y.~Zhang, M.~K. Kalra, F.~Lin, Y.~Chen, P.~Liao, J.~Zhou, and G.Wang.
\newblock Low-dose ct with a residual encoderdecoder convolutional neural
  network.
\newblock {\em IEEE Transactions on Medical Imaging}, 36(12):2524–--253,
  2017.

\bibitem{Jin2017}
K.~H. Jin, M.~T. McCann, E.~Froustey, and M.~Unser.
\newblock Deep convolutional neural network for inverse problems in imaging.
\newblock {\em IEEE Transactions on Image Processing}, 26(9):4509–--4522,
  2017.

\bibitem{Ye2018}
D.~H. Ye, G.~T. Buzzard, M.~Ruby, and C.~A. Bouman.
\newblock Deep back projection for sparse-view ct reconstruction.
\newblock In {\em IEEE Global Conference on Signal and Information Processing
  (GlobalSIP)}, pages 1--5, 2018.

\bibitem{Gupta2018}
H.~Gupta, K.~H. Jin, H.~Q. Nguyen, M.~T. McCann, and M.~Unser.
\newblock Cnn-based projected gradient descent for consistent ct image
  reconstruction.
\newblock {\em IEEE Transactions on Medical Imaging}, 37(6):1440–--1453,
  2018.

\bibitem{Anirudh2018}
R.~Anirudh, H.~Kim, J.~J. Thiagarajan, K.~A. Mohan, K.~Champley, and T.~Bremer.
\newblock Lose the views: Limited angle ct reconstruction via implicit sinogram
  completion.
\newblock In {\em 2018 IEEE/CVF Conference on Computer Vision and Pattern
  Recognition}, pages 6343–--635, 2018.

\bibitem{Wang2004}
Z.~Wang, A.C. Bovik, H.R. Sheikh, and E.P. Simoncelli.
\newblock Image quality assessment: from error visibility to structural
  similarity.
\newblock {\em IEEE Trans. Imag. Proc.}, 13(4):600--612, 2004.

\bibitem{NCI_Lung}
{N}ational~{C}ancer {I}nstitute.
\newblock {D}ata {S}cience {B}owl 2017 - {C}hest {CT} dataset for lung nodule
  detection.
\newblock Website, 2017.
\newblock available at
  \url{https://www.kaggle.com/c/data-science-bowl-2017/overview/description}.

\bibitem{LTT}
K.~M. Champley.
\newblock Livermore tomography tools (ltt) technical manual.
\newblock Tech. Rep. LLNL-SM-687016, LLNL, Livermore, CA, USA, 2016.

\end{thebibliography}
}

\end{document}